\documentstyle[12pt,a4]{article}
\begin{document}

\newcommand{\V}{{\bf V}}
\newcommand{\W}{{\bf W}}
\newcommand{\Na}{{\bf N}}
\newcommand{\si}{\sigma}
\newcommand{\pr}{\prime}
\newcommand{\br}[1]{|#1\rangle}
\newcommand{\be}{\begin{equation}}
\newcommand{\ee}{\end{equation}}
\newcommand{\bea}{\begin{eqnarray}}
\newcommand{\ena}{\end{eqnarray}}
\newcommand{\beas}{\begin{eqnarray*}}
\newcommand{\enas}{\end{eqnarray*}}
\newcommand{\vs}[1]{\rule[- #1 mm]{0mm}{#1 mm}}
\newcommand{\hs}[1]{\hspace{#1 mm}}
\newcommand{\sm}[2]{\frac{\mbox{\footnotesize #1}\vs{-2}}
                   {\vs{-2}\mbox{\footnotesize #2}}}
\newcommand{\half}{\frac{1}{2}}
\newcommand{\shalf}{\sm{1}{2}}
\newcommand{\slq}{U_q sl_2}
\newcommand{\hsl}{U_q\widehat{sl}_2}
\newcommand{\hg}{U_q\hat{g}}
\newcommand{\g}{U_qg}
\newcommand{\hr}{\hat{R}}
\newcommand{\vv}{v^{(1)}_0 \otimes v^{(2)}_0}
\newcommand{\vvv}{v^{(2)}_0 \otimes v^{(1)}_0}
\newcommand{\D}{\Delta}
\newcommand{\Sp}{{\bf S}}
\newcommand{\ot}{\otimes}
\newcommand{\up}{\uparrow}
\newcommand{\dow}{\downarrow}
\newcommand{\wn}{w_0^n}
\newcommand{\wnm}{w_m^n}
\newcommand{\nn}{\nonumber}
\newcommand{\la}{\lambda}
\newcommand{\Ml}{M_\lambda}
\newcommand{\hot}{\hat{\otimes}}
\newcommand{\insi}{\stackrel{\displaystyle <}{\sim}}
\newcommand{\susi}{\stackrel{\displaystyle >}{\sim}}
\newcommand{\op}{\oplus}
\renewcommand{\H}{{\cal H}}
\newcommand{\R}{\hat{R}}
\newcommand{\Pro}{{\cal Q}}
\newcommand{\Proj}{{\cal P}}
\newcommand{\ab}[1]{{\rm{\bf #1}}}
\renewcommand{\dim}{{\rm dim}}
\newcommand{\id}{{\rm id}}
\newcommand{\gr}[2]{ {( #1,#2)} }
\newcommand{\re}[1]{(\ref{#1})}
\newcommand{\r}{\rangle}
\newcommand{\lgl}{\langle}

\hfill NBI-HE-99-38

\vspace{12pt}
 
\begin{center} 
{\large \bf
Fermionisation of the Spin-$S$ Uimin-Lai-Sutherland Model: Generalisation
of the Supersymmetric t-J Model to  Spin-$S$}\footnote{This 
work was supported in part by INTAS grant 96-524} 

\vspace{36pt}

J.Ambj\o rn\footnote{e-mail:{\sl ambjorn@nbivms.nbi.dk}}

\vspace{6pt}
{\sl Niels Bohr Institute}

{\sl Blegdamsvej 17, Copenhagen, Denmark}

\vspace{24pt}

D.Karakhanyan\footnote{e-mail:{\sl karakhan@lx2.yerphi.am}},
M.Mirumyan\footnote{e-mail:{\sl ...@lx2.yerphi.am}}

\vspace{6pt}
{\sl Yerevan Physics Institute,}

{\sl Br.Alikhanian st.2, 375036, Yerevan, Armenia}

\vspace{24pt}

A.Sedrakyan\footnote{e-mail:{\sl sedrak@alf.nbi.dk}; permanent
address: Yerevan Physics Institute},

\vspace{12pt}

{\sl Niels Bohr Institute}

{\sl Blegdamsvej 17, Copenhagen,  Denmark}

\vspace{36pt}
September 1999


\vfill
{\bf Abstract} 
\end{center}
 
The $spin 1$ Uimin-Lai-Sutherland (ULS) isotropic chain model is 
expressed in 
terms of fermions and
the equivalence of the fermionic representation 
to the supersymmetric $t-J$ model is established
directly at the level of Hamiltonians.The spin-S ULS model is fermionized
and the Hamiltonian of the corresponding generalisation of the $t-J$ 
model is written down.

\newpage

\section{Introduction}
\indent

The low dimensional $(d= 0,1,2)$ integrable models, for a long time a 
subject of interest in the mathematical physics, are now attracting much
attention also in connection with problems in condensed matter
physics. The development of nanostructure technology nowadays allows us
to consider the problems in zero(quantum dots), one(quantum wires, 
organic polymers) and two(hall effect, high temperature superconductivity)
spatial dimensions.

The well known one dimensional Heisenberg model of a nearest-neighbour
interacting chain  of ${1 \over 2}$ spins, 
\begin{equation}
\label{H1}
H=\sum\limits_{i=1}^NH_{ii+1}=\sum\limits_{i=1}^N
\vec{\sigma_i}\vec{\sigma}_{i+1}, \;\;\;\qquad N+1=1,  
\end{equation}
was solved in 1931 by Bethe \cite{Bethe} using what is now 
called the Bethe Ansatz. During the last 25 years this method has been 
developed into what is now called the 
Quantum Inverse Scattering Method(QISM)  
\cite{Bax, F} (or the Algebraic Bethe Ansatz(ABA)). The
method allows one
to find a complete set of eigenvalues and eigenstates of the Hamiltonian
of a one dimensional integrable model.

The origin of integrability is a set of equations, first derived
in the article by Yang  \cite{YY}, and which, after the works of 
Baxter \cite{Bax}, appeared to be equivalent to existence  
of an infinite number conservation laws in the model, and called 
Yang-Baxter Equations (YBE) as a consequence.

However, the ABA technique was mainly developed in the context of the $XXZ$
spin chain model and its application to models formulated
originally in fermionic operator language was complicated. For example,
in the Hubbard model, solved in the very beginning by a Coordinate
Bethe Ansatz (CBA) method \cite{Lieb}, the implementation of ABA
to find a full solution was made only recently \cite{RM} after
remarkable step done by Shastry \cite{S1,S2,O}. He used the Jordan-
Wigner transformation to pass from fermionic to spin language,
in order to apply QISM, formulated in matrix form.

Recently the interest of formulating
spin chain models in terms 
of Fermi operators by use of Jordan-Wigner transformation 
\cite{Maas, Wad}, and by use some new method developed in \cite{HS,AHS,GM}, 
has increased. It appeared that the explicit construction of integrable
models and application of QISM in terms of Fermi operators becomes easy.
Because the Fock space of fermions is finite dimensional the 
investigation of the complete set of eigenstates and eigenvalues in 
this language becomes relatively simple.

It is well known that the trivial generalisation of the Heisenberg
model to higher spins is nonintegrable and gap-full \cite{H}. The first
integrable model for the spin-$S$ was established by G.Uimin \cite{U}
and later by J.K.Lai \cite{L} and B.Sutherland \cite{S}. They have 
considered the Hamiltonian
\begin{equation}
\label{ULS}
H =J \sum\limits_{i=1}^N
(Id_{i,i+1} + P_{i,i+1}), \qquad \qquad \qquad N+1=1,  
\end{equation}
where $P_{i,i+1}$ is the operator permuting spins at the sites $i$
and $i+1$ and $Id_{i,i+1}$ is the identity operator, 
shown its integrability and found a solution by the 
Coordinate Bethe Ansatz. Another integrable model for arbitrary
spin-$S$ was constructed in the articles \cite{Bab,T}.
As it appeared, the integrable Hamiltonian is a polynomial 
$Q_{2S}$ of
degree $2 S$ of the product of nearest-neighbour
spins $\vec{S}_i \vec{S}_{i+1}$
\begin{equation}
\label{BM}
H_S=J \sum\limits_{i=1}^N
Q_{2S}(\vec{S}_i\vec{S}_{i+1}) \;\;\;\qquad N+1=1,  
\end{equation}
where
\begin{eqnarray}
\label{Q1}
Q_{2S}(x)=-\sum\limits_{j=1}^{2S}(\sum\limits_{k=1}^{j})
\prod\limits_{\stackrel{l=0}{l\ne j}}^{2S}
{x-x_l \over x_j-x_l}, \nonumber\\
x_l={1 \over 2}[l(l+1)-2S(S+1)]
\end{eqnarray}
and has a critical point, near which the model is equivalent to
Wess-Zimino-Witten-Novikov (WZWN) model with corresponding
coupling constant $k=2S$ \cite{Af}.
However in the article \cite{Bab} only $2^L$ (L-is the length of
the chain) states from all $3^L$ amount of states were found together
with corresponding Bethe equations for them.

The full set of $SU(2)$ invariant, integrable, isotropic quantum spin
chain models for spins $S \leq 6$ was found in \cite{Ken}.

In this article we will use the technique developed in \cite{HS,AHS,GM},
which is alternative to the Jordan-Wigner transformation,
to fermionize the integrable spin-1 ULS isotropic chain model and
find its Hamiltonian in terms of fermionic fields, which will
demonstrate its equivalence 
to integrable supersymmetric (SUSY) $t-J$ model \cite{Kor}.
After that we generalise this construction to spin-$S$ and carry out 
the fermionization
of the spin-$S$ ULS chain model. The resulting Hamiltonian can be 
considered as 
spin-$S$ analogy for the $t-J$ model.

The article is organised as follows. In the Section 2 we describe
briefly the fermionization scheme for integrable models and QISM.

The fermionization of the spin-1 Uimin-Lai-Sutherland (ULS) model will
be carried out in the Section 3. The Hamiltonian will be written
in Fermi terms and coincidence with SUSY $t-J$ model will be
shown.

In the section 4 we will construct the fermionic representation
of the $SU(2)$ algebra for spins $(2S+1) \leq 8$ and apply that
construction to find the $R$-matrix and corresponding Hamiltonian
in terms of fermions in section 5.


\section{Yang-Baxter equation and structure of the quantum spaces of the
integrable model.}

 The key of integrability of the model is the Yang-Baxter equation
(YBE), which implies some restrictions on $R_{aj}$-matrix - the
constituent of the transfer matrix $T(u)$, defined as
\begin{equation}
T\left( u\right) =tr_a\prod\limits_{j=1}^NR_{aj}\left( u\right) .
\label{1.1}
\end{equation}
The YBE ensures the necessary and sufficient
conditions of commutativity of the transfer-matrices for different values 
of the
spectral parameter $u$, which has a physical meaning of the rapidity of
pseudoparticles of the model
\begin{equation}
\label{1.11}
[T\left( u\right),T\left( v\right)]=0.
\end{equation} 
The Hamiltonian of the model can be obtained by logarithmic derivative
of the transfer matrix at zero spectral parameter
\begin{equation}
\label{1.12}
H=-\frac{\partial \ln T\left( u\right)}{\partial u}\mid_{u=0}.
\end{equation} 

By definition $R_{aj}$ acts as a intertwining operator on the 
space of direct product of
the so called auxiliary $V_{a}\left( v\right) $ and quantum $V_{j}\left(
u\right) $ spaces 
\begin{equation}
R_{aj}\left( u,v\right):\quad V_{a}(u)\otimes V_{j}(v)\rightarrow
V_{j}(v)\otimes V_{a}(u)  \label{1.2}
\end{equation}
and can be represented graphically as
\vspace{-5mm}
\begin{center}
{\small {\ \setlength{\unitlength}{5mm} 
\begin{picture}(10,15)
\put(3.9,4.5){$V_j(v)$}
\put(0,8){\vector(1,0){7}}
\put(0.5,8.5){$V_a(u)$}
\put(3.5,10.6){\vector(0,1){0.7}}
\put(3,10.8){\line(-1,-1){0.2}}
\put(3,10.8){\line(1,-1){0.2}}
\multiput(3.5,5.4)(0,1){6}{\line(0,-1){0.7}}
\put(-4,7.85){$R_{aj}(u,v)=$} 
\put(0,3){\shortstack{Fig.1.}}
\end{picture}} }
\end{center}

\vspace{-5mm}

The spaces $V_a(u)$ and $V_j(v)$ with spectral parameters $u$ and $v$ are
irreducible representations of the affine quantum group $U_q\widehat{g},$
which is the symmetry group of the integrable model under consideration.
Provided that the states $\mid a\rangle $ $\in V_a$ and $\mid j\rangle \in
V_j$ form a basis for the spaces $V_a$ and $V_j$, following \cite{HS,AHS}
we can represent the action of the operator $R_{aj}$ as follows 
\begin{equation}
R_{aj}\mid j\rangle \otimes \mid a\rangle =R_{a^{\prime }j^{\prime
}}^{aj}\mid a^{\prime }\rangle \otimes \mid j^{\prime }\rangle , 
\label{1.3}
\end{equation}
where the summation is over the repeating indices $a^{\prime}$ and
$j^{\prime}$ (but not over $a$ and $j$).

By introducing the operators 
\begin{equation}
X_{a^{\prime }}^a=\mid a^{\prime }\rangle \langle a\mid; \qquad
X_{j^{\prime }}^j=\mid j^{\prime }\rangle \langle j\mid  
\label{1.4}
\end{equation}
in the graded spaces $V_a$ and $V_j$ correspondingly, one can 
easily rewrite (\ref{1.3}) as follows 
\begin{equation}
R_{aj}=R_{aj}\mid j\rangle \mid a\rangle \langle a\mid \langle j\mid
=R_{a^{\prime }j^{\prime }}^{aj}\mid a^{\prime }\rangle
\mid j^{\prime }\rangle \langle a\mid \langle j\mid =\left(
-1\right) ^{p(a)p(j^{\prime })}R^{a^{\prime }j^{\prime }}_{aj}X_{a^{\prime
}}^a X_{j^{\prime }}^j  
\label{1.5}
\end{equation}
where the sign factor takes into account the possible grading of the states $\mid
a\rangle $ and $\mid j\rangle ,$ $p(a)$ and $p(j)$ denote the corresponding
parities and the summation over the repeating indices. It is easy
to see from the expression (\ref{1.4}) that the operators $X_a^{a \prime}$ 
have the projection property
\begin{equation}
\label{1.41}
X_a^b X_{a^{\prime }}^{b^{\prime}}= \delta^b_{a^{\prime}}X_a^{b^{\prime}}
\end{equation}

In the conventional form the YBE looks like 
\begin{equation}
R_{a^{\prime }b^{\prime }}^{ab}(u,v)R_{a^{\prime \prime }j^{\prime
}}^{a^{\prime }j}(u,w)R_{b^{\prime \prime }j^{\prime \prime }}^{b^{\prime
}j^{\prime }}(v,w)=R_{b^{\prime }j^{\prime }}^{bj}(v,w)R_{a^{\prime
}j^{\prime \prime }}^{aj^{\prime }}(u,w)R_{a^{\prime \prime }b^{\prime
\prime }}^{a^{\prime }b^{\prime }}(u,v).  
\label{1.6}
\end{equation}
By use of equations (\ref{1.5}) and (\ref{1.41})
the matrix-valued YBE eq. (\ref{1.6}) one can easily be
transformed to the following 
operator valued equation
\begin{equation}
R_{a b}(u,v)R_{aj}(u,w)R_{bj}(v,w)=R_{bj}(v,w)R_{aj}(u,w)R_{a b}(u,v).  
\label{1.7}
\end{equation}

Using the graphical representation of the $R$-operator presented in Fig.1,
one can draw the YB equation graphically as follows

\unitlength .85mm
\hspace{-5mm}
\begin{picture}(140,50)
\put(12,15){\line(1,0){13}}
\put(12,30){\line(1,0){13}}
\put(25,15){\line(1,1){15}}
\put(25,30){\line(1,-1){15}}
\multiput(51,35)(0,-4){7}{\line(0,-1){2.8}}
\put(40,15){\vector(1,0){25}}
\put(40,30){\vector(1,0){25}}
\put(51,35){\vector(0,1){4}}
\put(10,10){$V_a(v)$}
\put(10,25){$V_a(u)$}
\put(52,33){$V_j(w)$}
\put(71.5,21.5){=}
\put(122,15){\vector(1,0){13}}
\put(122,30){\vector(1,0){13}}
\put(107,15){\line(1,1){15}}
\put(107,30){\line(1,-1){15}}
\put(107,15){\line(-1,0){25}}
\put(107,30){\line(-1,0){25}}
\multiput(96,35)(0,-4){7}{\line(0,-1){2.8}}
\put(96,35){\vector(0,1){4}}
\put(80,10){$V_a(v)$}
\put(80,25){$V_a(u)$}
\put(97,33){$V_j(w)$}
\put(68,-5){\shortstack{Fig.2}}
\end{picture}

\vspace{15mm}

An alternative form of the YBE above is the commutativity of the $R$%
-operator (\ref{1.5}) with the co-product $\Delta $ of the corresponding
quantum group $U_q\widehat{g}$ \cite{HS,AHS}.

Let us introduce now the $L$-operator by taking matrix element of 
the $R$-operator 
(\ref{1.5}) in the auxiliary space $V_a$ with the basic vectors $\mid a\rangle 
$ \cite{GM} 
\begin{equation}
\label{1.8}
L_{a^{\prime }}^a\left( u,v\right) =\langle a^{\prime }\mid R(u,v)\mid
a\rangle =\left( -1\right) ^{p(a)p(j)}R_{a^{\prime }j^{\prime }}^{aj}\left(
u,v\right) X_j^{j^{\prime }} 
\end{equation}
Then, following \cite{GM}, it is easy to show that YBE takes the form 
\begin{equation}
\check R_{ab}^{a^{\prime}b^{\prime}}
\left( L\otimes _sL\right)_{a^{\prime}b^{\prime}}^{a^{\prime \prime}
b^{\prime \prime}} 
=\left( L\otimes _sL\right)_{a b}^{a^{\prime} b^{\prime}}
\check R_{a^{\prime}b^{\prime}}^{a^{\prime \prime}
b^{\prime \prime}}  
\label{1.9}
\end{equation}
where $\check R_{cd}^{ab}= R^{ba}_{cd}$ and graded direct
product is defined as follows
\begin{equation}
\label{SP}
(A\otimes _sB)_{bd}^{ac}=\left( -1\right) ^{(p(a)+p(b))p(d)}A_b^aB_d^c,
\end{equation}
which is conventional in the graded Quantum Inverse Scattering Method 
\cite{KS}.

\section{The $spin$ $1$ ULS-chain in terms of fermions}
\indent

Now we would like to apply the approach, described above, to fermionize
the $spin$ $1$ integrable isotropic ULS chain with the Hamiltonian
(\ref{ULS}).

To begin with we should realize the $spin$ $1$ algebra in terms of 
fermions. 
At least two sort of fermions are needed in order to express three
basic states $\mid +\rangle$, $\mid 0\rangle$, $\mid -\rangle$ 
of the $spin 1$
particle
with the z component of the spin equal to  $1, 0, -1$ 
respectively. 
Let us define $c^{+}_{\sigma}, \;\;c_{\sigma},\;\;\; 
\sigma=\uparrow,\downarrow$ as a  
creation
-annihilation operators of fermions with the up and down spins 
correspondingly, together with their Fock space 
$\mid 0\rangle, \mid \sigma\rangle$.

The generators of $spin1$ representation of the $su(2)$ algebra 
can be
built by use of this Fermi operators as follows: 
\begin{eqnarray}
\label{3.1}
S^{+} &=&\left( 1-n_{\uparrow}\right)c_{\downarrow}
+(1-n_{\downarrow})c^{+}_{\uparrow},  \nonumber \\
S^{-} &=&\left( 1-n_{\uparrow}\right)c^{+}_{\downarrow}
+(1-n_{\downarrow}) c_{\uparrow}, \\
S_z &=&n_{\uparrow} - n_{\downarrow}, \nonumber
\end{eqnarray}
where $n_{\sigma}$ is the fermion number operators.

By this definition of $su(2)$ generators, the states 
with definite third projection are realized through fermionic Fock 
space as follows
\begin{equation}
\label{FS}
\mid +\rangle \equiv \mid\uparrow\rangle \mid 0\rangle,
\;\;\;\;
\mid 0\rangle \equiv \mid 0 \rangle \mid 0\rangle,\;\;\;\;
\mid -\rangle \equiv \mid 0 \rangle \mid \downarrow \rangle.
\end{equation}
Indeed, easy to
check that
$S_z|+\rangle =|+\rangle ,$ $S_z|0\rangle
=0,$ $S_z|-\rangle =-|-\rangle ,$ $S_{+}|+\rangle =0,$ $S_{+}|0\rangle
=|+\rangle ,$ $S_{+}|-\rangle =|0\rangle ,$ $S_{-}|+\rangle =|0\rangle ,$ $
S_{-}|0\rangle =|-\rangle ,$ $S_{-}|-\rangle =0$. 

As it is obvious
from formulas (\ref{FS}), we have constructed a graded space 
with the following
parities for the basic vectors
\begin{equation}
\label{Par}
p(\mid +\rangle)=p(\mid -\rangle)= 1,
\;\;\;\;
p(\mid 0\rangle)= 0.
\end{equation}

The sum of the square of spin generators, which is the Casimir operator 
\begin{equation}
\label{3.2}
S^{+}S^{-}+S^{-}S^{+}+S_z^2\equiv {\cal{C}} =2(1-
n_{\uparrow}n_{\downarrow}), 
\qquad S_a{\cal{C}} =2S_a={\cal{C}} S_a  
\end{equation}
determines the projection operator 
\begin{equation}
\label{3.3}
\Pi =1-n_{\uparrow} n_{\downarrow},\qquad 
\left( 1-n_{\uparrow} n_{\downarrow}\right)^2
=1-n_{\uparrow} n_{\downarrow}, 
\end{equation}
which maps 4-dimensional space of direct product of two $spin1/2$ spaces
onto 3-dimensional subspace $spin1$. ${1 \over 2}{\cal C}$ acts
as an identity operator on eigenstates of the $S_z$:
\begin{equation}
\label{3.4}
\frac 12{\cal{C}} \mid m\rangle =\mid m\rangle \qquad m =+,-,0 , 
\end{equation}
 and annihilates the fourth
state 
\begin{equation}
\label{3.5}
{\cal{C}} \mid \uparrow\rangle \mid \downarrow \rangle =0,
\end{equation}
which corresponds to scalar state in the expansion of direct product
of two half spins.

In the $spin 1$ case the operator $X$ takes the form 
\begin{eqnarray}
\label{3.6}
X_{i m}^k \equiv |m\rangle _{i}\langle k|_i &=&\left( 
\begin{array}{lll}
|-\rangle \langle -| &\qquad |-\rangle \langle 0| 
&\qquad |- \rangle \langle +| \\ 
|0\rangle \langle -| &\qquad |0\rangle \langle 0| 
&\qquad |0\rangle \langle +| \\ 
|+\rangle \langle -| &\qquad |+\rangle \langle 0| 
&\qquad |+\rangle \langle +|
\end{array}
\right) _i \nonumber\\
&=&\left( 
\begin{array}{lll}
(1-n_{\uparrow}) n_{\downarrow} &(1-n_{\uparrow})c^{+}_{\downarrow} 
&c^{+}_{\downarrow} c_{\uparrow} \\ 
(1-n_{\uparrow})c_{\downarrow}& (1-n_{\uparrow})(1- n_{\downarrow}) 
&c_{\uparrow}(1-n_{\downarrow}) \\ 
c^{+}_{\uparrow} c_{\downarrow}& c^{+}_{\uparrow}(1-n_{\downarrow}) 
& n_{\uparrow}(1- n_{\downarrow})
\end{array}
\right) _i , 
\end{eqnarray}
where $i$ denotes the chain site.
Trace of this operator is an identity operator due to the completeness of
set of the states. 

The operator $\Pi_{ij}$, which permutes the states between spaces 
$V_i$ and $V_j$ has the form
\begin{equation}
\label{2.3}
\Pi_{ij}= \sum\limits_{m,k}\mid m_j\rangle \mid k_i\rangle 
\langle k_j \mid \langle m_i \mid =
(-1)^{p(k)}X^m_{ik}X^k_{jm} 
\end{equation}
The sign in this expression reflects the grading of the spaces $V_i$

One can find the $R$-matrix of the $spin 1$ model in its bosonic form in
\cite{U, L, S}, but it can easily be guessed also from the expression of the
Hamiltonian (\ref{ULS}),
\begin{equation}
\label{3.81}
H= {1 \over 4}\sum\limits_{i=1}^{N}\left[\vec{S}_i\vec{S}_{i+1}+
(\vec{S}_i\vec{S}_{i+1})^2 \right],
\end{equation}
which can be represented also as Unity + Permutation operators 
in the 3-dimensional $spin 1$ space. It has a following form
\begin{equation}
\label{3.8}
\check R_{ms}^{kq}(u)=\delta _m^k\delta _s^q+ u \;
\delta _s^k\delta _m^q, 
\qquad m,k,s,q =+,0,-,
\end{equation}
where first term is unit and second term bosonic permutation matrices.

The L-operator of the model, defined in fermionic representation by the
formula (\ref{1.8}), takes the form
\negthinspace 
\begin{eqnarray}
 \label{3.7}
L_i(u)&=&\left( 
\begin{array}{lll}
u-\left( 1-n_{\uparrow}\right) \left( 1-n_{\downarrow}\right) 
& \left( 1-n_{\uparrow}\right) c^{+}_{\downarrow} 
& -c^{+}_{\uparrow} c^{+}_{\downarrow} \\ 
\left( 1-n_{\uparrow}\right) c_{\downarrow} 
& u+\left( 1-n_{\uparrow}\right)n_{\downarrow} 
& c^{+}_{\uparrow}n_{\downarrow} \\ 
c_{\uparrow}^{+} c_{\downarrow} & c_{\uparrow} n_{\downarrow} 
& u-n_{\uparrow} n_{\downarrow}
\end{array} \right) _i \nonumber\\
&=& \left( 
\begin{array}{lll}
u+\frac 12\left( S_z^2-S_z\right) & -S_{+}S_z & S_{+}^2 \\ 
-S_zS_{-} & u-1+S_z^2 & S_zS_{+} \\ 
-S_{-}^2 & S_{-}S_z & u+\frac 12\left( S_z^2+S_z\right)
\end{array}
\right) _i  
\end{eqnarray}
and the Yang-Baxter equation  (\ref{1.9}) takes form 
\begin{eqnarray}
\label{3.9}
\!\! &&L_m^k\left( u\right) L_s^q\left( v\right) +(u-v)\left( -1\right)
^{p(m)p(q)+p(k)p(q)+p(m)p(k) }L_s^k\left( u\right) 
L_m^q\left(v\right)  \nonumber\\
&=&L_m^k\left( v\right) L_s^q\left( u\right) +(u-v)\left( -1\right)
^{p(q)p(s)+p(m)p(q)+p(s)p(m) }L_m^q\left( v\right) 
L_s^k\left(u\right) .   
\end{eqnarray}

>From the equations above one can be easily read off the invariance 
of the YBE under the transformations 
\begin{equation}
L_m^k\left( u\right) \rightarrow L_m^{\prime k}\left( u\right)
=M_q^k L_p^q\left( u\right) N_m^p  \label{3.10}
\end{equation}
where $M$ and $N$ are arbitrary numerical matrices of the 
same grading as $ L\left( u\right) $.

According to the prescription of the previous section, by use of 
formulas
(\ref{1.5}), (\ref{3.8}) and the definition (\ref{1.12})
as a logarithmic derivative of
transfer matrix, it is easy to find 
the following
expression for the Hamiltonian via graded permutation 
(\ref{2.3}) and identity operators 

\begin{equation}
\label{3.11}
H=\sum\limits_{i=1}^NH_{ii+1}=\sum\limits_{i=1}^N\left(
I_{ii+1}+\Pi_{ii+1}\right), 
\end{equation}
due to the fact that ${R}_{ik}\left( 0\right) =P_{ik}$. 

In terms of Fermi operators this Hamiltonian
can be represented in very simple way
as 
\begin{equation}
\label{3.13}
H_{i j}=\Delta _i\Delta_j\left( I_{i j}+{\cal{P}}_{i j,\uparrow}
{\cal{P}}_{i j,\downarrow}
\right) \Delta _i\Delta _j ,
\end{equation}
where $\Delta _i$ is projection operator defined in (\ref{3.3}) and 
${\cal{P}}_{i j,\sigma} \equiv 1-(c_{i, \sigma}^{+}-c_{j, \sigma}^{+})
(c_{i, \sigma}-c_{j, \sigma}),\;\;$ 
$\sigma= \uparrow, \downarrow$ 
are the fermionic permutation operators.

It is now straightforward to recognise in the expression (\ref{3.13})
the Hamiltonian of supersymmetric $t-J$ \cite{AHS,GM,Kor,EK1} model.

The symmetry (\ref{3.10}) mentioned above is generated by 
the operator 
\begin{eqnarray}
\label{3.14}
Q_{ib}^a &=&\left( 
\begin{array}{lll}
\left( 1-n_{\uparrow}\right) n_{\downarrow} 
& \left( 1-n_{\uparrow}\right) c_{\downarrow}^{+} 
& c^{+}_{\downarrow} c_{\uparrow} \\ 
\left( 1-n_{\uparrow}\right)c_{\downarrow} 
& \left( 1-n_{\uparrow}\right) \left(1-n_{\downarrow}\right) 
& c_{\uparrow}\left(1 -n_{\downarrow}\right) \\ 
c^{+}_{\uparrow} c_{\downarrow} & c^{+}_{\uparrow}\left(1- 
n_{\downarrow}\right) 
& n_{\downarrow} n_{\downarrow}
\end{array}
\right) _i , \nonumber\\
Q_b^a&=&\sum\limits_{i=1}^NQ_{ib}^a,\qquad\qquad
\left[ H,Q_b^a\right] =0.
\end{eqnarray}
In fact $Q_{i}$ coincides with the operator $X_i$ (\ref{3.6}).
Being defined as global quantity along the chain, the  matrix $Q$ is 
inert under
permutation of the site indices, i.e. it commute with the local
Hamiltonians $H_{ii+1}$.

In the unrestricted Fock space of spin-up and spin-down fermions
one can construct the $X$-operator (\ref{1.4}), which will be a graded
direct product of two $X$'s in the one-fermion space
\begin{eqnarray}
\label{3.141}
(X)_{ a b}^{a^{\prime} b^{\prime}}=(X_{\uparrow} \otimes_s 
X_{\downarrow})_{a b}^{a^{\prime} b^{\prime}}, \qquad\qquad
\qquad a,b=0,1.
\end{eqnarray}
It is easy to see, that the $spin$-1 $X$-operator, defined as
in eq.(\ref{3.6}), can be obtained from the expression (\ref{3.141})
 by deleting the row and column, corresponding
to the exclusion of state 
$\mid \uparrow \rangle \mid \downarrow \rangle $. 


\section{Spin-S Representation of the SU(2) algebra in the Fock space of 
Fermi fields}
\indent   

In order to describe the $2 S +1$ dimensional space of the $spin-S$
representation by fermions we will consider the Fock space of $r$
copies of the fermionic creation-annihilation operators $c_{\mu}^{+},
c_{\mu}, \mu=1,2...r$.
\begin{equation}
\label{T1}
{\cal{V}}_r=\bigotimes_{\mu =1}^{r} V_{\mu},
\end{equation}
where $V_{\mu}$ is the Fock space of $\mu$-type fermions.
The dimension of this space is $2^r$ and hence the necessary condition
 to have a enough degrees of freedom for the $spin-S$ states is 
$(2 S+1) \leq 2^r$.

As usual let us mark the basic elements of the fermionic Fock space
by their filling numbers $n_\mu =0,1 ; \mu =1,...r$
\bea
\label{F1}
|n_1,...n_r\rangle &=& |n_1\rangle \cdot|n_2\rangle\cdot \cdot
\cdot |n_r\rangle =
(c_1^{+})^{n_1} (c_2^{+})^{n_2}\cdot \cdot \cdot
(c_r^{+})^{n_r}|0\rangle \\
|n_1,...n_r\rangle &\in& {\cal{V}}_r \nonumber
\ena

It is necessary to order this basic vectors by integers which can be
achieved, for example, by use of binary system of numbers. But let us
define $m$ as a map of the set of sequences $(n_1,n_2,...n_r)$ onto 
integers ${1,2,.. 2^r}$
\be
\label{F2}
m:{(n_1,n_2,...n_r)}\rightarrow{1,2...2^r} 
\ee
Then it is clear, that
\be
\label{F22}
{\cal V}_r = \bigotimes_{m=1}^{2^r} W_m
\ee
where $W_m$ is the space defined by the basic vector 
$|n_1,n_2,...n_r\rangle$.

There is a natural grading in the Fock space of fermions. Zero and
one particle states in $V_\mu$ define a graded homogeneous
subspaces  $V_{\mu}= V_{0 \mu} \oplus V_{1\mu}$ with the parity
$p$ as a function $V_{j\mu}\rightarrow Z_2$
\be
\label{F3}
p(|n_{\mu}\r)= n_{\mu},\qquad |n_{\mu}\r \in V_{n_{\mu},\mu},
\qquad n_{\mu}=0,1
\ee
Grading of the tensor product space ${\cal V}_r =
\ot_{\mu=1}^{r} V_{\mu}$ is
defined by parities of its constituents, namely, by parities of the
basic elements of ${\cal V}_r$, as follows
\be
\label{F4}
p(|n_1,...n_r\r)=\sum\limits_{\mu=1}^{r} p(|n_{\mu}\r).
\ee

Now, following section 2 let us define the operators
\bea
\label{F5}
X_m^{m^{\prime}}&=&X_{n_1,...n_r}^{n_1^{\pr},...n_r^{\pr}}=
|n_1,...n_r\r \langle n_r^{\pr},...n_1^{\pr}|
=(-1)^{(p(n_r)+p(n_r^{\pr}))\sum\limits_{l \ne r}p(n_l^{\pr})}\cdot
 \cdot\nn\\
&&\cdot \cdot \cdot
(-1)^{(p(n_{2})+p(n_{2}^{\pr}))p(n_1^{\pr})}X_{n_1}^{n_1^{\pr}} \cdot
\cdot \cdot
X_{n_r}^{n_r^{\pr}},
\ena
where 
\be
\label{F6}
X_{n_{\mu}}^{n_{\mu}^{\pr}}= |n_{\mu}\r \lgl n_{\mu}^{\pr}|,\qquad
\mu=1,...r
\ee
is the corresponding $X$-operator in the Fock space $V_{\mu}$ 
of the $\mu$-type fermions. Operators $X_{n_{\mu}}^{n_{\mu}^{\pr}}$
can be considered as a matrix $X_{\mu}$ with the operator valued
entities of the form
\bea
\label{F7}
X_{\mu}= \left( 
\begin{array}{ll}
1-n_{\mu}&c_{\mu}\\
c_{\mu}^{+}&n_{\mu}
\end{array}\right).
\ena

By use of definition (\ref{SP}) the expression (\ref{F5}) can
be represented as a graded direct product of the operators
$X_{n_{\mu}}^{n_{\mu}^{\pr}}$
\be
\label{F71}
X_{n_1,...n_r}^{n_1^{\pr},...n_r^{\pr}}=\left(X_1 \ot_s \cdot
\cdot \cdot \ot_s X_r\right)_{n_1,...n_r}^{n_1^{\pr},...n_r^{\pr}}
\ee

It is now meaningful to define also the operators which projects
an arbitrary state from the representation space ${\cal V}_r$ on to
the particular one-dimensional subspace defined by the state
$|n_1,...n_r\r$. Obviously it is
\bea
\label{F8}
P_{m(n_1,...n_r)}&=&P_{n_1,...n_r}=|n_1,...n_r\r \lgl n_r,...n_1|\nn\\
P_{m(n_1,...n_r)}^2&=&P_{m(n_1,...n_r)}
\ena
Therefore the operator
\be
\label{F9}
\D_{m(n_1,...n_r)}=1 -P_{n_1,...n_r}
\ee
excludes the one dimensional subspace, defined by the basic vector
$|n_1,...n_r\r$, from the whole Fock space ${\cal V}_r$, reducing
the dimension by one. Hence, in order to construct a $(2 S +1)$
dimensional representation space ${\cal V}_{2S+1}$ for the $spin-S$
 we should act by corresponding amount of projectors $\D_m$ of the
type (\ref{F9}) on the whole $2^r$-dimensional space ${\cal V}_r$
\be
\label{F10}
{\cal V}_{2S+1}=\D_1 \cdot \cdot \cdot \D_{2^r-(2S+1)} {\cal V}_r.
\ee 

The definition (\ref{F10}) means that any operator $A$ which
acts on the space ${\cal V}_r$, will have a form
\be
\label{F11}
A_S =\D_1\cdot \cdot \cdot \D_{2^r-(2S+1)} A 
\D_1\cdot \cdot \cdot \D_{2^r-(2S+1)}
\ee
on the  space ${\cal V}_{(2S+1)}$.

Again, let's now numerate the remaining basic directions in 
${\cal V}_(2S+1)$ by $l(n_1,...n_r)=1,2...(2S+1)$ and denote
the corresponding $X$-operator as
\be
\label{F12}
X_l^{l^{\pr}}=\D_1 \cdot \cdot \cdot \D_{2^r-(2S+1)} 
X_{n_1,...n_r}^{n_1^{\pr},...n_r^{\pr}}
\D_1 \cdot \cdot \cdot \D_{2^r-(2S+1)}.
\ee 

Then, as it is easy to check by use of
formula (\ref{1.41}) and some simple algebraic calculations, that
the operators $S^B, B=1,2,3$ defined by
\be
\label{F13}
S^B=\left(S^B \right)_l^{l^{\pr}}\cdot X_{l^{\pr}}^l,\qquad 
l,l^{\pr}=1,...(2S+1),
\ee
 fulfill the commutation relations of the $SU(2)$-algebra
generators. In (\ref{F13}) $\left(S^B \right)_l^{l^{\pr}}$ are
the ordinary number valued matrix elements of the $spin-S$ generators
of the $SU(2)$-algebra, the only nonzero elements of which are
\bea
\label{F14}
\left(S^+\right)_l^{l-1}&=&\left(S^+\right)_M^{M-1}=
\sqrt{(S+M)(S-M+1)},\qquad S^-=\left(S^+\right)^+,\nn\\
\left(S_3\right)_l^{l}&=&\left(S_3\right)_M^{M}=M,\qquad\qquad
\qquad \qquad M=-S+l-1.
\ena

>From now on we will consider the case $r=3$ and the spins 
$S \leq (2^3-1)/2=7/2$ for simplicity.

For the convenience let us numerate the even elements of basic
vectors $|n_1,...n_r\r$ as $\alpha,\beta =1,2,3,4$ in the
following way
\be
\label{QQ1}
|0,1,1\r \equiv 1,\qquad |1,0,1\r \equiv 2,\qquad 
|1,1,0\r \equiv 3, \qquad |0,0,0\r \equiv 4.
\ee
and the odd elements $a,b=5,6,7,8$  as
\be
\label{QQ2}
|1,0,0\r \equiv 5,\qquad |0,1,0\r \equiv 6,\qquad 
|0,0,1\r \equiv 7, \qquad |1,1,1\r \equiv 8,
\ee
Hence we have
\be
\label{F15}
{\cal V}_3= \bigoplus_{\alpha=1}^{4}W_{\alpha}
\bigoplus_{a=5}^{8}W_{a}=V_{odd}\op V_{even}
\ee

Now by use of definition (\ref{F5}) and (\ref{F7}) one
can find the following expression for the 
$X_m^{m^{\pr}}$-operator $(m,m^{\pr}=1,...8)$ in the
unrestricted space ${\cal V}_3$
\bea
\label{F16}
X_m^{m^{\pr}}=\left(
\begin{array}{ll}
X_{\alpha}^{{\alpha}^{\pr}}&X_{\alpha}^{a^{\pr}}\\
X_{a}^{{\alpha}^{\pr}}&X_{a}^{{a}^{\pr}}
\end{array}\right),
\ena
where
\bea
\label{F161}
X_{\alpha}^{{\alpha}^{\pr}}=\left(
\begin{array}{llll}
(1-n_1)n_2 n_3&-c_1 c_2^+ n_3&c_1 n_2 c_3^+&
(1-n_1)c_2^+ c_3^+\\
c_1^+ c_2 n_3&n_1 (1-n_2)n_3&-n_1 c_2 c_3^+&
c_1^+ (1-n_2)c_3^+\\
-c_1^+ n_2 c_3&n_1 c_2^+ c_3& n_1 n_2 (1-n_3)&
c_1^+ c_2^+(1-n_3)\\
-(1-n_1)c_2 c_3&-c_1 (1-n_2)c_3&  -c_1 c_2 (1-n_3)&
(1-n_1)(1-n_2)(1-n_3)
\end{array}\right),
\ena
\bea
\label{F162}
X_{a}^{{a}^{\pr}}=\left(
\begin{array}{llll}
n_1(1-n_2)(1-n_3)&-c_1^+ c_2(1-n_3)&c_1^+(1- n_2)c_3
&-n_1 c_2 c_3\\
-c_1 c_2^+(1-n_3)&(1-n_1) n_2(1-n_3)&(1-n_1)c_2^+ c_3
&c_1 n_2 c_3\\
-c_1(1-n_2)c_3^+&-(1-n_1)c_2 c_3^+&(1- n_1)(1-n_2)n_3
&-c_1 c_2 n_3\\
n_1 c_2^+c_3^+&-c_1^+n_2 c_3^+&  c_1^+ c_2^+ n_3&
n_1 n_2 n_3,
\end{array}\right),
\ena
\bea
\label{F163}
X_{a}^{\alpha}=\left(
\begin{array}{llll}
-c_1^+c_2 c_3&-n_1(1-n_2)c_3&-n_1 c_2(1-n_3)&
c_1^+(1-n_2)(1-n_3)\\
-(1-n_1)n_2c_3&c_1 c_2^+c_3&c_1n_2(1-n_3)&
(1-n_1)c_2^+(1-n_3)\\
(1-n_1)c_2n_3&c_1(1-n_2)n_3&-c_1 c_2c_3^+&
(1-n_1)(1-n_2)c_3^+\\
c_1^+ n_2 n_3&-n_1 c_2^+ n_3&n_1 n_2 c_3^+&
c_1^+ c_2^+ c_3^+,
\end{array}\right),
\ena
and
\be
\label{F164}
X_{\alpha}^{a}=\left(X_{a}^{\alpha}\right)^+.
\ee

It is easy to see from the expressions (\ref{F161}) and (\ref{F162}) that
$X_{\alpha}^{{\alpha}^{\pr}}$ can be obtained from the
$X_{a}^{{a}^{\pr}}$ by particle-hole transformation of all fermions
in a following way
\be
\label{F17}
c_1 \leftrightarrow c_1^+, \qquad c_2 \leftrightarrow  -c_2^+,
\qquad c_3 \leftrightarrow c_3^+
\ee

\section{Fermionic representation of the spin-S ULS model:
Spin-S generalisation of the supersymmetric t-J model}
\indent

In this section we will fermionize the Hamiltonian of the ULS
model defined by the formula (\ref{ULS}) for spins 
$S \leq {7 \over 2}$, which can be described by three Fermi
fields. The procedure is based on the approach described in the 
section 2 and is an alternative to the usual Jordan-Wigner 
transformation.
First we consider spin ${7 \over 2}$.

We start with the $R$-matrix of the ULS model
\begin{equation}
\label{R1}
\check R_{ms}^{kq}(u)=\delta _m^k\delta _s^q+ u \;
\delta _s^k\delta _m^q, 
\qquad m,k,s,q =1,2,...8.
\end{equation}
and in the same way as for $spin$-1 case described in the section
3, we obtain the Hamiltonian
\be
\label{R2}
H_{7 \over 2}=\sum\limits_{i=1}^{N}H_{i,i+1}=
\sum\limits_{i=1}^{N}\left(I_{i,i+1}+\Pi_{i,i+1}\right)=
\sum\limits_{i=1}^{N}\left(I_{i,i+1}+
\prod\limits_{\mu=1}^{3}{\cal P}_{i,i+1;\mu}\right)
\ee

Our goal is now to rewrite this expression of the Hamiltonian in a way
which is similar to $spin 1$ $ t-J$ model.

According to formula (\ref{F11}) the Hamiltonian for spins
less that ${7 \over 2}$ will be
\be
\label{R3}
H_S = \D_1 \cdots \D_{8-(2S+1)}H_{7 \over 2}\D_{8-(2S+1)} \cdots \D_1,
\ee
where the projector $\D_1 \cdots \D_{8-(2S+1)}$ cuts off some 
subspace
from the full 8-dimensional Fock space of representation ${\cal V}_3$,
reducing the degrees of freedom to $(2S+1)$. The different choices
of the cutted subspaces (projectors $\D$) corresponds to isomorphic
representations in a mathematical sense and simply means 
particle-hole transformations for some of three fermions from a 
physical point of view.

Let us now expand the expression (\ref{R2}) for the Hamiltonian
$H_{i,i+1}$ with the signs, as it should be according to
eq.(\ref{1.5}) and substitute the expressions for the $X$-operators
(\ref{F161} -\ref{F164}) . Then after some algebra and by use
of definitions (\ref{F8}) of the projectors we 
obtain
\bea
\label{R4}
H_{j,j+1}= H^{1,hop}_{j,j+1}+H^{2,hop}_{j,j+1}+
{\bar H}^{2,hop}_{j,j+1}+H^{diag}_{j,j+1}+H^3_{j,j+1}
-{\bar H}^{3}_{j,j+1}.
\ena
The different terms in this Hamiltonian is defined as follows.
The terms in $H^{1,hop}_{j,j+1}$ are the one and three particle hopping
terms
\bea
\label{R5}
H^{1,hop}_{j,j+1}&=&(X_j)_{m}^{{\alpha}}
(X_{j+1})^{m}_{{\alpha}}-
(X_j)^{m}_{{\alpha}}(X_{j+1})_{m}^{{\alpha}}\nn\\
&=&c_{j,1}^+\left(P_{j,4}P_{j+1,4}+P_{j,6}P_{j+1,6}+P_{j,7}P_{j+1,7}+
P_{j,1}P_{j+1,1}\right)c_{j+1,1}\nn\\
&+&c_{j,2}^+\left(P_{j,5}P_{j+1,5}+P_{j,4}P_{j+1,4}+P_{j,7}P_{j+1,7}+
P_{j,2}P_{j+1,2}\right)c_{j+1,2}\nn\\
&+&c_{j,3}^+\left(P_{j,5}P_{j+1,5}+P_{j,6}P_{j+1,6}+P_{j,4}P_{j+1,4}+
P_{j,3}P_{j+1,3}\right)c_{j+1,3}\nn\\
&-&\left(c_{j,1}^+c_{j,2}c_{j,3}c_{j+1,1}c_{j+1,2}^+c_{j+1,3}^+ -
c_{j,1}c_{j,2}^+c_{j,3}c_{j+1,1}^+c_{j+1,2}c_{j+1,3}^+\right.\nn\\
&-&\left. c_{j,1}c_{j,2}c_{j,3}^+c_{j+1,1}^+c_{j+1,2}^+c_{j+1,3} -
c_{j,1}^+c_{j,2}^+c_{j,3}^+c_{j+1,1}c_{j+1,2}c_{j+1,3} -h.c.\right)
\ena

We have extracted the terms, which contains $\alpha =4$ and $m=8$
indices into separate $H^{2,hop}_{j,j+1}$ and ${\bar H}^{2,hop}_{j,j+1}$
hopping terms, correspondingly. 
\bea
\label{R6}
H^{2,hop}_{j,j+1}&=&(X_j)_4^{\alpha}
(X_{j+1})_{\alpha}^4 + (X_j)^4_{\alpha}(X_{j+1})^{\alpha}_4\\
&=&-P_{j,8}\left(c_{j,2}c_{j,3}c_{j+1,2}^+c_{j+1,3}^+ +
c_{j,1}c_{j,3}c_{j+1,1}^+c_{j+1,3}^+ 
+c_{j,1}c_{j,2}c_{j+1,1}^+c_{j+1,2}^+\right)P_{j+1,8}\nn
\ena
As it is clear from the formulas (\ref{F163}) and (\ref{F164})
the ${\bar H}^{2,hop}_{j,j+1}$ term can be obtained from the
$H^{2,hop}_{j,j+1}$ by particle hole transformation (\ref{F17})
These terms are responsible for the hopping of the Fermi pairs.

$H^{diag}$ is the diagonal term containing coulomb interaction
of fermions
\bea
\label{R7}
H^{diag}&=&(X_{j})_{\alpha}^{\alpha}(X_{j+1})_{a}^{a}+
(X_{j})_4^4(X_{j+1})_{\alpha}^{\alpha}-
(X_{j})_{8}^{8}(X_{j+1})_{a}^{a} +(j \leftrightarrow j+1)\nn\\
&=&\left(1-P_{j,1}-P_{j,2}-P_{j,3}\right)\left(P_{j+1,1}+P_{j+1,2}
+P_{j+1,3}+P_{j+1,4}\right)\nn\\
&-&P_{j,8}\left(P_{j+1,5}+P_{j+1,6}
+P_{j+1,7}+P_{j+1,8}\right)+(j \leftrightarrow j+1).
\ena

The last two terms in (\ref{R4}) are analogous to the spin-spin
interaction term in the ordinary $t-J$ model and,
 as in previous case, ${\bar H}^{3}_{j,j+1}$
can be obtained from the $H^3_{j,j+1}$ by the particle hole 
transformation (\ref{F17})
\bea
\label{R8}
H^3_{j,j+1}&=&-(X_j)_{\bar{a}}^{\bar{a}}
(X_{j+1})^{\bar{b}}_{\bar{b}}-
(X_j)_{\bar{a}}^{\bar{b}}
(X_{j+1})^{\bar{a}}_{\bar{b}}=\qquad \qquad \bar{a},\bar{b}=1,2,3\nn\\
&=& \hat{S}^B_j \hat{S}^B_{j+1} +(\hat{S}^B_j \hat{S}^B_{j+1})^2,
\qquad \qquad \qquad \qquad \qquad B=1,2,3
\ena
where
\bea
\label{R9}
\hat{S}^B_j &=& (S^B)_{\bar{b}}^{\bar{a}} X_{\bar{a}}^{\bar{b}}=
\psi^+_{\bar{b}}(S^B)_{\bar{b}}^{\bar{a}} \psi_{\bar{a}},\nn\\
\psi_{\bar{a}} &=&c_{\bar{a}}(1-n_1)(1-n_2)(1-n_3),\qquad \qquad
\bar{a}=1,2,3.
\ena
and $(S^B)_{\bar{b}}^{\bar{a}}$ is the $spin-1$ representation of the
SU(2)-algebra.

This expression can be obtained easily from 
\be
\label{R10}
X_{\bar{a}}^{\bar{b}}=\psi^{+\bar{b}}\psi_{\bar{a}}.
\ee
which follows from formula (\ref{F161}).
By appropriate an choice of projectors $\D_{\sigma}, \,\,
\sigma=1,...8-(2S+1)$ we can simplify the expression for the
Hamiltonian, for example by deleting the terms $H^{{2,hop}}$ and
${\bar H}^{{2},hop}$.  

\section{Acknowledgements}
The authors acknowledges INTAS grant-0524 for financial support.  
J.A. and A.S thanks MaPhySto -- Centre for Mathematical Physics and
Stochastics, funded by a grant from The Danish National Research
Foundation -- for support.


\end{document}